\documentstyle[preprint,aps,tighten]{revtex}

\input epsf

\begin{document}

\draft

\preprint{UGI-98-17}

\title{Interpretation and resolution of pinch singularities \\  
in non-equilibrium quantum field theory}
\author{{\bf Carsten Greiner}\footnotemark[1]
{\bf and Stefan Leupold}\footnotemark[3]}
\address{Institut f\"ur Theoretische Physik, Universit\"at Giessen,
\\
D-35392 Giessen, Germany}

\footnotetext[1]{e-mail address: carsten.greiner@theo.physik.uni-giessen.de}
\footnotetext[3]{e-mail address: stefan.k.leupold@theo.physik.uni-giessen.de}

\date{April 1998}

\maketitle

\begin{abstract}
Ill-defined pinch singularities arising
in a perturbative expansion in out of equilibrium quantum field
theory have a natural analogue to standard scattering theory.
We explicitly demonstrate that the occurrence of such terms is directly related
to Fermi's golden rule known from elementary scattering theory and is thus
of no mystery. 
We further argue that within the process of thermalization of a plasma
one has to resum such contributions to all orders as the process itself
is of non-perturbative nature. In this way the resummed propagators obtain
a finite width. Within the Markov approximation of kinetic theory
the actual phase space distribution
at a given time of the evolution enters explicitly.
\end{abstract}

Non-equilibrium many-body theory or quantum field theory has become
a major topic of research for describing various transport processes
in nuclear physics, in cosmological particle physics or more generally
in quantum dissipative systems. A very powerful diagrammatic tool is given
by the `Schwinger-Keldysh' \cite{Sc61,BM63,Ke64} or `closed time path' (CTP) 
technique by means of non-equilibrium Green's functions
for describing a quantum system
also {\em beyond} thermal equilibrium \cite{Ch85}. For an equilibrium situation
this technique is equivalent to the real time description
of finite temperature field theory \cite{Mi69,La87,Bel96}.

Employing the diagrammatic CTP rules potential `pinch singularities'
might arise in strictly perturbative expressions. As an example
we consider a scalar field theory.
A typical contribution arising in a perturbative expansion
takes e.g.~the form
\begin{equation}
\label{pinch2}
D_0^{\rm ret}(\vec{ p}, p_0 ) \Sigma_0(\vec{ p}, p_0 )
D_0^{\rm av}(\vec{ p}, p_0 )
\, \, \, .
\end{equation}
Here $\Sigma_0$ describes some physical (perturbative) quantity 
(e.g.~a self energy insertion);
$D_0^{\rm ret}$ and $D_0^{\rm av}$ denote the free retarded and advanced
propagator, respectively. As $D_0^{\rm ret}$ contains a pole
at $p_0 = \pm E_p -i\epsilon $ and $D_0^{\rm av}$ a pole
at $p_0 = \pm E_p +i\epsilon $ the product of both in the above expression
is ill-defined, if $\Sigma_0(\vec{ p}, p_0 = E_p=\sqrt{m^2+\vec{p}^2})$ 
does {\em not} 
vanish onshell. Transforming such an expression back into a time representation,
the contour has to pass between this pair of two infinitely close poles.

It was observed and proven by Landsman and van Weert that such ill-defined
terms cancel each other in each order in perturbation theory, if the
system stays at {\em thermal} equilibrium \cite{La87}. Their arguments, however,
rely solely on the KMS boundary conditions of the free propagators
and self energy insertions, so that
they do not apply for systems {\em out} of equilibrium.
This severe problem arising for systems out of equilibrium was first
raised by Altherr and Seibert \cite{AS94}. Indeed, it was speculated there 
that the CTP formalism might not be adequate for describing non-equilibrium
systems at all.
In a subsequent paper, Altherr
\cite{A95} tried to `cure' this problem by hand by introducing a finite width for
the `unperturbed' free CTP propagator $D_0$ so that the expressions
are at least well-defined in a mathematical sense. 
Within his modified perturbative approach,
he also showed
that seemingly higher order diagrams do contribute to a lower order in the
coupling constant, as some of the higher order diagrams
involving pinch terms will receive factors of the form
$1/\Gamma ^n, \, n\geq 1$ reducing substantially the overall power in the
coupling constant.
In his particular case Altherr investigated the dynamically generated
effective mass (the `tadpole' contribution) within standard $\phi^4-$theory.
(For the hard modes the onshell damping $\Gamma $ is of the order of $o(g^4 T)$.)
Therefore he concluded that power counting arguments might in fact
be much less trivial for systems out of equilibrium.
We will come back to his observation below.

In a recent work \cite{GL98} we have discussed in detail that
modes or quasi-particles become thermally populated
by a non-perturbative Langevin like interplay
between noise and dissipative terms entering the non-equilibrium quantum
transport equations. In the process of thermalization the
{\em full} propagators necessarily must acquire some finite
width (due to collisions or more generally due to damping). 
Plasmons behave as  `nonshell' modes \cite{La88}.
Strictly speaking, the evolution of a non-equilibrium
system towards equilibrium is always {\em non-perturbative}.
We will come back to this interpretation in more detail below.

First, however, we will elaborate on the 
{\em physical reason} for the occurrence of pinch singularities in
a strictly perturbative expansion, when an interacting system is prepared
with some non-equilibrium occupation of the particles. As a motivation
we were inspired by the idea that in principle the Schwinger-Keldysh formalism
is also adequate to describe simple scattering processes where
e.g.~only two initial particles are prepared at some fixed
momentum states in the past. Hence, the perturbative scheme of the
Schwinger-Keldysh formalism should give the same results as elementary
scattering theory. 

\begin{figure}
\centerline{\epsfxsize=5cm \epsfbox{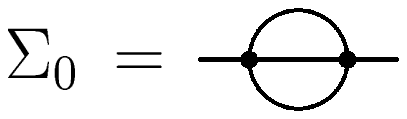}}

\caption{Lowest order self energy term in $\phi^4$-theory 
which contributes to the pinch problem (sunset diagram).}
\label{fig:sunset}
\end{figure}

\begin{figure}
\centerline{\epsfxsize=7cm \epsfbox{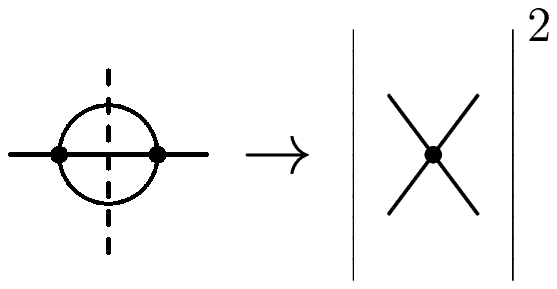}}

\caption{Imaginary part of the sunset diagram which can be identified with
a scattering amplitude.}
\label{fig:scat}
\end{figure}

To set the stage we start with some formulae
and manipulations already presented in \cite{AS94}.
We follow the notation of \cite{GL98}.
For simplicity we consider in the following a weakly interacting scalar 
$\phi^4$-theory.
The initial state in the far past
(assuming a homogeneous and stationary system)
is prepared by specifying the momentum occupation number $\tilde{n}(\vec{p})$ of the
(initially non interacting) onshell particles. Note that this occupation number
depends only on the three momentum $\vec{p}$. (If specified
with a thermal equilibrium distribution at some given temperature, $\tilde{n}$
would be replaced by the onshell Bose distribution
$n_B(E_p=\sqrt{m^2+\vec{p}^2})$.) The occupation number $\tilde{n}(\vec{p})$ enters 
the (free) propagator
\begin{equation}
D_0^< (p)\, = \, - 2\pi i
\, {\rm sgn}(p_0) \,\delta (p^2-m^2)
\left[
\Theta(p_0) \tilde{n} (\vec{p})
\,
- \, \Theta(-p_0) (1+\tilde{n} (\vec{p})) \right]
\label{D0<noneq}
\end{equation}
In addition, we note the form of the free retarded and advanced propagator:
\begin{eqnarray}
D_0^{\rm ret/av}(p) &=& \frac{1}{p^2-m^2 \pm i \epsilon \, {\rm sgn}(p_0)} \, .
\label{D0ret}
\end{eqnarray}
To calculate perturbative corrections to the propagators
we apply the Langreth-Wilkins rules \cite{LW72}
which are quite well-known within the context of the Schwinger-Keldysh formalism:
\begin{eqnarray}
D^{\rm ret} & = & D_0^{\rm ret} \, + \,
D_0^{\rm ret}\Sigma_0 ^{\rm ret} D_0 ^{\rm ret} \, + \ldots  
=: D_0^{\rm ret} \, + \, \Delta D^{\rm ret}  \,,
\label{DSDret1} \\
D^{\rm av} & = & D_0^{\rm av} \,
+ \, D_0^{\rm av}\Sigma_0 ^{\rm av} D^{\rm av} \, + \ldots
=: D_0^{\rm av} \, + \, \Delta D^{\rm av}  \,,
\label{DSDav1} \\
D^{<} & = & D_0^{<} \, + \,
D_0^{\rm ret}\Sigma_0^{\rm ret}D_{0}^{<} \, + \,
D_0^{\rm ret}\Sigma_0^{<}D_0^{\rm av} \, + \,
D_{0}^{<} \Sigma_0^{\rm av}D_0^{\rm av} \, + \ldots
=: D_0^{<} \, + \, \Delta D^{<}  \,,
\label{DSD<1}
\end{eqnarray}
where the dots denote multiple self energy insertions which we will not consider 
for the moment. Here the retarded and advanced
self energies are given by the Fourier transforms of (cf.~e.g.~\cite{GL98})
\begin{eqnarray}
\label{eq:defsigret}
\Sigma^{\rm ret}(x_1,x_2) & := & 
\Theta(t_1-t_2)\,\left[ 
\Sigma^>(x_1,x_2) - \Sigma^<(x_1,x_2) 
\right]   \, , \\
\label{eq:defsigav}
\Sigma^{\rm av}(x_1,x_2) & := & 
\Theta(t_2-t_1)\,\left[ 
\Sigma^<(x_1,x_2) - \Sigma^>(x_1,x_2) 
\right]  \, .
\end{eqnarray}
The self energies 
$\Sigma^>$ and $\Sigma^<$ are related as
$\Sigma^>(x_1,x_2) = \Sigma^<(x_2,x_1)$ in case of a scalar field theory.
The self energy insertion $\Sigma_0$ in a strictly perturbative expansion
is given by a convolution of the {\it initial} free propagators.
If the initial momentum distribution entering 
the propagator (\ref{D0<noneq}) is given by the Bose
equilibrium distribution,
the important relation
\begin{equation}
  \Sigma^>(p) = e^{p_0/T}\, \Sigma^<(p) \,,  \label{eq:heat}
\end{equation}
holds, which is nothing but the KMS boundary condition.
It is worth mentioning that our conventions are chosen such that  
$i \Sigma^< (p) $ is always real and non negative. In a transport
theory (see below) it can be interpreted as the production rate for modes
with the respective energy.
As a characteristic example we discuss in the following the 
`sunset' graph arising in scalar $\phi ^4$-theory.
This diagram is illustrated in fig.~\ref{fig:sunset}. 
We choose this particular graph
as an example since the self energies $\Sigma_0 ^{</>} (\vec{p},p_0=E_p)$ do not
vanish onshell for thermal distributions (see e.g.~\cite{EH95,CG97}).
This also holds
for any non-equilibrium distribution $\tilde{n}$
as long as the individual two-particle scattering contributions
are kinematically allowed. Within finite temperature field theory the imaginary
part of the self energy (`cut' diagram) taken onshell is 
connected to the scattering rate (as an illustration see 
fig.~\ref{fig:scat}).
On the other hand, there exist certain self energy insertions
like the so-called hard thermal loop self energy \cite{Car97}
or other one-loop diagrams \cite{Da98} which vanish on-shell
due to simple kinematical constraints and thus
do not cause any pinch problem.

By inspecting (\ref{DSDret1}-\ref{DSD<1}) more closely one finds
that the perturbative corrections $\Delta D^{\rm ret/av}$ to the free 
retarded/advanced propagator are free of any pinch singularities
as the emerging poles are all located at the same side of the contour. 
We note in passing that this also holds
for multiple self energy insertions in (\ref{DSDret1},\ref{DSDav1}) 
(see e.g.~\cite{GL98}).
In contrast, all three contributions to $\Delta D^{<} $ are ill-defined.
Using the identity
\begin{equation}
\label{identity}
\pi {\rm sgn}(p_0) \,\delta (p^2-m^2) \, = \,
\frac{i}{2} \left( D_0^{\rm ret}(p) - D_0^{\rm av}(p) \right)
\end{equation}
together with (\ref{D0<noneq}) we can further
manipulate the three contributions of $\Delta D^{<} $
by employing the Fourier transforms of the definitions (\ref{eq:defsigret})
and (\ref{eq:defsigav}). We find
\begin{equation}
  \label{eq:regpinch}
\Delta D^{<} (p) = \Delta D^{<}_{\rm reg} (p)  + \Delta D^{<}_{\rm pinch} (p)
\end{equation}
with a regular part, 
\begin{eqnarray}
\Delta D^{<}_{\rm reg} (p) & = & \left[ \Theta(p_0) \tilde{n} (\vec{p}) \,
- \, \Theta(-p_0) (1+\tilde{n} (\vec{p})) \right]   
\nonumber \\ && 
  \label{eq:reg}
\times \left( D_0^{\rm ret}(p) \Sigma_0 ^{\rm ret}(p) D_0^{\rm ret}(p) \, - \,
D_0^{\rm av}(p) \Sigma_0 ^{\rm av}(p) D_0^{\rm av}(p) \right)   \,,
\end{eqnarray}
and the part carrying the pinch singularities, 
\begin{eqnarray}
\Delta D^{<}_{\rm pinch} (p) & = &
D_0^{\rm ret}(p)
\left[
\Theta (p_0) \left(
(1+\tilde{n} (\vec{p})) \Sigma_0 ^<(p) - \tilde{n} (\vec{p}) \Sigma_0 ^>(p)
\right)  \right.   \nonumber \\ 
&& \phantom{mmm} \left. \, + \, \Theta (-p_0) \left(
(1+\tilde{n} (\vec{p})) \Sigma_0 ^>(p) - \tilde{n} (\vec{p}) \Sigma_0 ^<(p)
\right)
\right]
D_0^{\rm av} (p) \, .  
\label{DSD<2} 
\end{eqnarray}
The last expression is ill-defined, 
if the terms in the square brackets do not vanish onshell as already pointed out
in \cite{AS94}. The expression in the square brackets
is familiar from standard kinetic theory (see e.g.~\cite{AS94,GL98}):
Apart from a trivial factor one can interpret
\begin{equation}
\label{netrate}
\Gamma_{\rm eff} (\vec{p}) \, := \,
\frac{1}{2 E_p}
\left. \left[
(1+\tilde{n} (\vec{p})) i \Sigma_0 ^>(p) - \tilde{n} (\vec{p}) i \Sigma_0 ^<(p)
\right]
\, \vphantom{\int} \right\vert_{p_0=E_p}
\end{equation}
as the net effective rate for the change of the occupation number per time.
For an equilibrium situation the occupation number
is given by the Bose distribution and the self energy insertions fulfill
the KMS condition (\ref{eq:heat}). Hence, for the equilibrium case
the whole bracket exactly vanishes and no pinch singularities emerge.
In contrast, this is not the case for a general non-equilibrium configuration
\cite{AS94}.

To shed first some light on the physical interpretation of this
ill-defined expression one has to ask for observables which are affected
by this singularity. Within standard scattering theory one would think about
the probability for a particle of some initial momentum state to be
scattered into another momentum state. Therefore we ask, 
how the occupation number $\tilde{n}$ has changed after a long time. 
The occupation number for the out-states can be readily extracted
from $D^{<}$ by means of the formula (for a derivation see \cite{GL98})
\begin{eqnarray}
\label{partnumb}
\lefteqn{n(\vec{p},t\rightarrow \infty )^{\rm (out)} = 
\langle a^{\dagger\, {\rm (out)} }_{\vec{p}} a_{\vec{p}}^{\rm (out)} \rangle  }
\nonumber \\
\label{partnumb1}
& = &
\left. \left( \frac{E_p}{2} +
\frac{1}{2E_p} \frac{\partial }{\partial t} \frac{\partial }{\partial t'}
+ \frac{i}{2}  ( \frac{\partial }{\partial t} -\frac{\partial }{\partial t'} )
\right) \frac{1}{V}
\int d^3x \int d^3y \, e^{i\vec{p} \vec{x}} e^{-i\vec{p} \vec{y}}
\left( i D^<(\vec{y},t;\vec{x},t') \right) \, \right| _{t'=t}
 \\
\label{partnumb2}
& \stackrel{t \rightarrow \infty }{=}&
\int \frac{dp_0}{2\pi }
\left( \frac{E_p}{2} + \frac{p_0^2}{2E_p} + p_0 \right)
\left( i D^<(\vec{p},p_0) \right)
\, \, \, . 
\end{eqnarray}
When inserting (\ref{eq:regpinch}) one finds by contour integration 
that $\Delta D^{<}_{\rm reg}$ only yields a finite contribution. The same holds true
for the $\Theta(-p_0)$-term in (\ref{DSD<2}) since the 'particle projector' 
$\left( \frac{E_p}{2} + \frac{p_0^2}{2E_p} + p_0 \right)$ vanishes on the 
antiparticle mass shell. 
However, the $\Theta(p_0)$-term of the ill-defined expression 
$\Delta D^{<}_{\rm pinch}$ gives rise to the following infinite expression
\begin{equation}
\label{delparta}
\Delta n(\vec{p})^{\rm (out)} \,  = \,
\Gamma _{\rm eff} (\vec{p}) \cdot 2\pi \delta (0) + \mbox{finite contributions.}
\end{equation}
From analogy to the standard derivation of Fermi's golden rule
in elementary quantum scattering theory one is immediately tempted
to interpret this $\delta(0)$ singularity as the elapsed scattering
time $T\rightarrow \infty $. Indeed, this interpretation
has very recently been conjectured by Niegawa in \cite{Ni97},
where he was also elaborating on the issue of pinch singularities
in non-equilibrium quantum field theory.
His major point, however, has been to interpret the {\em infinite}
shift $\Delta n(\vec{p}) $ as a renormalization in the number density.
We think, however, that this latter interpretation further obscures the problem
instead of uncovering the physical processes which are at the bottom of the
pinch problem. 

To demonstrate that the pinch singularities indeed appear 
as a result of Fermi's golden rule in scattering theory
we now assume that the interaction is switched on at a time
$t=-T/2$ and switched off at $t=T/2$, i.e.~we replace
\begin{equation}
\label{model}
\Sigma_0^{</>}(x_1,x_2) \rightarrow 
\bar{\Sigma }_0^{</>}(x_1,x_2) := 
\Theta({\textstyle {T \over 2}} -t_1) \, \Theta({\textstyle \frac{T}{2}}-t_2) \,
\Sigma_0^{</>}(x_1,x_2) \,
\Theta(t_1+{\textstyle \frac{T}{2}}) \, \Theta(t_2+{\textstyle \frac{T}{2}})
\end{equation}
and assume that the duration time $T$ is large but finite.
This procedure regulates the pinch singularity 
to a finite value.
As a first step we again extract the pinch term from (\ref{DSD<1}), now working in
the representation of three-momentum  and time:
\begin{eqnarray}
\lefteqn{\Delta D^<_{\rm pinch} (\vec{p},t,t')  }
\nonumber \\
&=& \int \frac{dp_{0(1)}}{2\pi } \frac{dp_{0(2)}}{2\pi } \frac{dp_{0(3)}}{2\pi }
\, e^{-ip_{0(1)}t} \, e^{ip_{0(3)}t'} \,
D_0^{\rm ret}(\vec{p}, p_{0(1)}) \,
D_0^{\rm av}(\vec{p}, p_{0(3)}) \,
\nonumber \\
&& \times 
\left[
\left(
\Theta(p_{0(1)}) \tilde{n} (\vec{p})
- \Theta(-p_{0(1)}) 
(1+\tilde{n} (\vec{p})) \right)\Sigma_0^{\rm av}(\vec{p},p_{0(2)})
\, + \, \Sigma_0^<(\vec{p},p_{0(2)}) \right.
\nonumber \\
&&
\phantom{mmm}
\left. \, - \,
\Sigma_0^{\rm ret}(\vec{p},p_{0(2)})
\left( \Theta(p_{0(3)}) \tilde{n} (\vec{p})
- \Theta(-p_{0(3)}) (1+\tilde{n} (\vec{p})) \right)
\right]
\nonumber \\
\label{D<pinch}
&& \times
\int\limits^{T/2}_{-T/2} d\bar{t} \, e^{i\bar{t}(p_{0(1)}-p_{0(2)})} \,
\int\limits^{T/2}_{-T/2} d\bar{t}' \, e^{i\bar{t}'(p_{0(2)}-p_{0(3)})}  \, \, .
\end{eqnarray}
As
\begin{equation}
\label{delta}
\int\limits^{T/2}_{-T/2} dt \, e^{it\Delta p} \, = \,
\frac{2}{\Delta p} \sin \left(\frac{T}{2} \Delta p \right) \,
\stackrel{T \rightarrow \infty }{\rightarrow } \,
2 \pi \delta (\Delta p) \, \, ,
\end{equation}
it becomes clear how the pinch singularity arises for $T\rightarrow \infty $.
Furthermore, if $T$ is already sufficiently large, we are safely allowed to 
approximate
$p_{0(1)} \approx p_{0(2)} \approx p_{0(3)}$ within the square bracket 
in (\ref{D<pinch}):
\begin{eqnarray}
\label{approx}
[ \ldots ]
&\approx &
\left[
\Theta (p_{0(2)}) \left(
(1+\tilde{n} (\vec{p})) \Sigma_0 ^<(\vec{p},p_{0(2)}) - \tilde{n} (\vec{p})
\Sigma_0 ^>(\vec{p}, p_{0(2)}) \right)
\, \right.
\\
&& \left. \, + \, \Theta (-p_{0(2)}) \left(
(1+\tilde{n} (\vec{p})) \Sigma_0 ^>(\vec{p},p_{0(2)}) - \tilde{n} (\vec{p})
\Sigma_0 ^<(\vec{p}, p_{0(2)}) \right)
\right] \, .
\nonumber
\end{eqnarray}
We proceed by calculating $\Delta n(\vec{p})^{\rm (out)}$
by means of (\ref{partnumb1}). For this we first take $t,t'> T/2$,
evaluate the $p_{0(1)}$- and $p_{0(3)}$-integration by standard
complex contour integration and then insert the emerging expression into
(\ref{partnumb1}). It results in
\begin{eqnarray}
\lefteqn{\Delta n(\vec{p})^{\rm (out)}_{\rm pinch} \, = \, 
(-i)
\left( \frac{E_p}{2} +
\frac{1}{2E_p} \frac{\partial }{\partial t} \frac{\partial }{\partial t'}
+ \frac{i}{2}  ( \frac{\partial }{\partial t} -\frac{\partial }{\partial t'} )
\right)  } \nonumber
\\
&&  \otimes
\int \frac{dp_{0(2)}}{2\pi } \,
\left[
\Theta (p_{0(2)}) \left(
(1+\tilde{n} (\vec{p})) \Sigma_0 ^<(\vec{p},p_{0(2)}) - \tilde{n} (\vec{p})
\Sigma_0 ^>(\vec{p}, p_{0(2)}) \right)
\, \right.  \nonumber 
\\
&& \phantom{mmmmm} \left. \, + \, \Theta (-p_{0(2)}) \left(
(1+\tilde{n} (\vec{p})) \Sigma_0 ^>(\vec{p},p_{0(2)}) - \tilde{n} (\vec{p})
\Sigma_0 ^<(\vec{p}, p_{0(2)}) \right)
\right] 
\nonumber
\\
&& \phantom{m} \times  \frac{1}{E_p} 
\left[ 
\frac{\sin(\frac{T}{2}(p_{0(2)}+E_p))}{p_{0(2)}+E_p} e^{iE_p t}
- 
\frac{\sin(\frac{T}{2}(p_{0(2)}-E_p))}{p_{0(2)}-E_p} e^{-iE_p t}
\right]
\nonumber
\\
&& \left. \phantom{m} \times  
\frac{1}{E_p} 
\left[ 
\frac{\sin(\frac{T}{2}(p_{0(2)}-E_p))}{p_{0(2)}-E_p} e^{iE_p t'}
- 
\frac{\sin(\frac{T}{2}(p_{0(2)}+E_p))}{p_{0(2)}+E_p} e^{-iE_p t'}
\right]  \,\, \right\vert_{\,t'=t}
\nonumber
\\[5mm]
\label{delpartb}
& \approx & \Gamma _{\rm eff} (\vec{p}) \,
\int \frac{dp_{0(2)}}{2\pi } \,
\frac{4}{(p_{0(2)}-E_p)^2} \, \sin ^2 \left(\frac{T}{2}(p_{0(2)}-E_p) \right)
\,  = \,
\Gamma _{\rm eff} (\vec{p}) \cdot T \, \, 
\end{eqnarray}
which is valid for large but finite $T$. 

Thus we have demonstrated the bridge
between the occurrence of pinch singularities within the context of the CTP
formalism and Fermi's golden rule in elementary quantum
scattering theory. The effective rate $\Gamma _{\rm eff}$ is therefore analogous
to the {\em transition probability per unit time}.
Indeed one can easily understand 
in physical terms that one has to expect such a singularity in perturbation theory:
Staying strictly within the first order contribution the particles remain
populated with the initially prepared non-equilibrium occupation number (since this
quantity enters the {\it free} propagator (\ref{D0<noneq})) and
scatter for an infinitely long time. Therefore,
the resulting shift $ \Delta n(\vec{p})^{\rm (out)}$
should scale with $\Gamma _{\rm eff} (\vec{p}) \cdot T $ with
$\Gamma _{\rm eff} (\vec{p})$ held fixed. We conclude that the occurrence
of pinch singularities appearing in perturbative contributions
within non-equilibrium quantum field theory is of no
mystery, but actually it has to appear because of a very intuitive reason:
the interaction time $T$ becomes infinite. However, looking at a Boltzmann equation
which describes the time evolution of the particle distribution function in
the semiclassical regime (see (\ref{eq:boltz}) below) one realizes that 
the occupation number does not stay constant during the 
dynamical evolution of the system, but will be changed on a timescale
of roughly $1/\Gamma $. 
The quasi-particles are not really asymptotic states.

Next, however, we will show how pinch 
singularities are formally cured by a resummation procedure. 
The onshell non-equilibrium effective rate $\Gamma_{\rm eff}$
can be visualized as being the net result of collisions between
the onshell particles. From standard thermal field theory one would thus
expect that the propagators will become dressed and supplemented
by a finite (collisional or more generally damping) width.
This represents already a non-perturbative effect which only
can be achieved by a resummation of Dyson-Schwinger type.
As a first attempt (proposed by Baier et al.~\cite{Ba97}), 
one might resum the full series of (\ref{DSDret1}-\ref{DSD<1})
using the self energy $\Sigma_0$ (recall that the latter is calculated from 
{\it free} propagators):
\begin{equation}
\label{DSeqa}
D =  D_0 \, + \, D_0 \Sigma_0 D_0 \, + \,
D_0 \Sigma_0 D_0 \Sigma_0 D_0 \, + \ldots
= D_0 \, + \, D_0 \Sigma_0 D    \,.
\end{equation}
With the definitions
$\Gamma _0(\vec{ p}, p_0) :=
\frac{i}{2 p_0 }[\Sigma _0^>(\vec{ p},p_0 ) - \Sigma _0^<(\vec{ p},p_0 ) ]$ and
${\rm Re}\Sigma _0 := {\rm Re}\Sigma _0^{\rm ret}  = {\rm Re}\Sigma _0^{\rm av} $
we end up with (cf.~e.g.~\cite{GL98})
\begin{eqnarray}
\label{Dret}
D^{\rm ret} & = & D_0^{\rm ret} \, + \, D_0^{\rm ret}\Sigma _0 ^{\rm ret}
D^{\rm ret} \, = \,
\frac{1}{p^2 - m^2 - {\rm Re}\Sigma_0 +ip_0 \Gamma _0} \, ,
\\
\label{Dav}
D^{\rm av} & = & D_0^{\rm av} \, + \, D_0^{\rm av}\Sigma _0 ^{\rm av}
D^{\rm av} \, = \,
\frac{1}{p^2 - m^2 - {\rm Re}\Sigma_0 - i p_0 \Gamma _0} \, ,
\\
\label{D<}
D^{<} & = & D^{\rm ret} \Sigma _0^< D^{\rm av} \, = \,
(-2i)
\frac{p_0 \Gamma _0}{(p^2-m^2-{\rm Re} \Sigma _0 )^2 +
p_0^2 \Gamma _0 ^2 }
\, \frac{\Sigma _0^< }{\Sigma _0^> - \Sigma _0 ^<} \, \, \, .
\end{eqnarray}
Hence the resummation of the series (\ref{DSD<1})
of ill-defined terms
results in a well-defined expression.
The quantity
\begin{eqnarray}
n_{\Sigma } (\vec{ p}, p_0 )\,  := \,
\frac{\Sigma _0^< }{\Sigma _0^> - \Sigma _0^<} 
\label{nsig}
\end{eqnarray}
appearing in (\ref{D<}) has to be interpreted as the `occupation number' demanded
by the self energy parts \cite{GL98}.
If the equilibrium KMS conditions (\ref{eq:heat}) apply for the self energy part,
then $ n_{\Sigma } (\vec{ p}, p_0 )$
$\stackrel{\mbox{KMS}}{\longrightarrow }$ $n_B(p_0 ) $
becomes just the Bose distribution function.
For a general non-equilibrium situation, however, this factor
deviates from the Bose distribution.
If the damping width is sufficiently small, 
i.e.~$\bar\Gamma , \, \bar\Sigma^>,\, \bar\Sigma^<$ 
are proportional to some power in the (small) coupling constant
$g$ (e.g.~$\sim g^4$ in case of the sunset graph depicted in fig.~\ref{fig:sunset})
the expression (\ref{D<}) results in
\begin{equation}
D^{\rm ret} \Sigma _0^< D^{\rm av} \,
\stackrel{g \rightarrow 0}{\longrightarrow } \,
-2\pi i \, {\rm sgn}(\omega)\,\delta (p^2-m^2) \, \lim_{g\rightarrow 0} 
n_{\Sigma } (\vec{ p}, p_0)   \, .
\label{DSD>4}
\end{equation}
When evaluating the occupation number for the out-states
by means of (\ref{partnumb2}) one accordingly will get 
\begin{equation}
\label{numpartc}
n(\vec{p})^{\rm (out)} \,  \approx \,
n_{\Sigma } (\vec{ p}, E_p) 
\end{equation}
which is free of any pathological behavior. 
The astonishing thing to note at this point is that in fact
the (initial) non-equilibrium distribution $\tilde{n}$
has been substituted by $n_{\Sigma }$ and, therefore, does not show up explicitly.
So the question is, {\em how} $\tilde n$ enters? 

Calculating $\Sigma_0$ on
a purely perturbative level the initial occupation number $\tilde n$ enters via
the free propagator (\ref{D0<noneq}). This however cannot be the whole truth
in a dynamically evolving system.
It is important to make sure that such a 
system is prepared at some {\it finite} initial time $t_0$.
(If $t_0$ would be taken as $t_0 \rightarrow - \infty$ the system would already
have reached equilibrium long time ago.)
Bedaque \cite{B95} already has noted that pinch singularities are in fact an 
artifact of the boundaries chosen at $t_0 \rightarrow -\infty $.
Time reversal symmetry is
explicitly broken, so that the propagators in principle
have to depend on
both time arguments explicitly before the system has reached a final
equilibrium configuration. Therefore the use of Fourier techniques (which in fact
has led to the pinch singularities in (\ref{DSD<2})) is highly dubious. 
The initial out of equilibrium distribution
$\tilde{n}(t_0)$ cannot stay constant during the evolution of the system
as it has to evolve towards the Bose distribution.
Hence there must exist contributions which attribute
to the temporal change of the distribution function.
As long as the system is not in equilibrium
(on  a time scale of roughly $1/\Gamma_0(\vec{p},E_p)$),
the propagator thus cannot be stationary.
In addition, the self energy parts
$\Sigma^<$ and $\Sigma^>$ do also evolve with time. Hence they should depend
on the {\em evolving} distribution function and not persistently on the
initial one, $\tilde{n}$, which enters $\Sigma _0$ in (\ref{DSeqa}).
Thus the resummation of (\ref{DSeqa}) does not cover all relevant contributions.
Speaking more technically, the self energy operators must also
be evaluated consistently by the fully dressed and temporally evolving
one-particle propagators.

\begin{figure}
\centerline{\epsfxsize=13cm \epsfbox{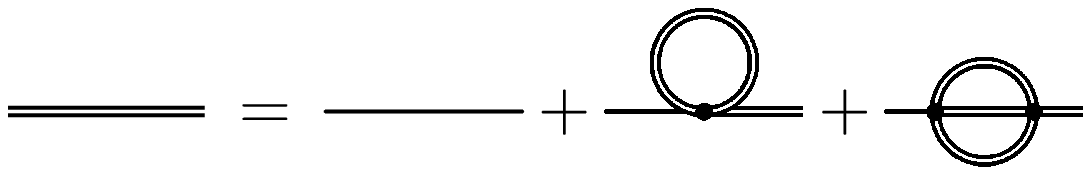}}

\caption{Dyson-Schwinger equation with fully dressed propagators (skeleton 
expansion).}
\label{fig:full}
\end{figure}

The solution to these demands is, of course, the description
of the system by means of appropriate (quantum) transport equations
\cite{KB,Ch85,GL98}. Graphically this is illustrated in fig.~\ref{fig:full}.
In addition to the sunset diagram we have
also included the mean field or Hartree diagram there
which in a perturbative scheme is the one which arises first. (It would,
however, not result in a pinch singularity so that we had discarded
it in our previous discussion.)
The difference to the resummation of (\ref{DSeqa}) is the fact
that the propagators entering into the self-energy operators are now
also the fully dressed ones.
Such a skeleton expansion of the self energies with including 
the dressed propagators
in the resummation is also familiar in standard quantum many-body theory
for strongly interacting systems \cite{FW}.

Unfortunately, the full quantum transport
equations are generally hard to solve and thus are not so much of practical use.
Yet one need not be that pessimistic. If the coupling is weak,
i.e.~the damping width is sufficiently small
compared to the quasi-particle energy (which
one typically assumes for many applications) one can take the Markov
approximation to obtain standard kinetic equations (for a derivation see 
e.g.~\cite{KB,Ch85,GL98}). For the situation illustrated in fig.~\ref{fig:full} 
one gets the standard form \cite{GL98}
\begin{eqnarray}
  \label{eq:boltz}
\lefteqn{\left( E_p \partial _t -\vec{p}\partial_{\vec{x}}
- \partial_{\vec{x}} m(\vec{x},t) \partial_{\vec{p}} \right)
f(\vec{x},t;\vec{p})  }
\\
&=&  {1\over 2} \left[
i \Sigma^<(\vec{x},t;\vec{p},E_p) \, (f(\vec{x},\vec{p},t)+1)
- i \Sigma^>(\vec{x},t;\vec{p},E_p) \, f(\vec{x},\vec{p},t)
\right] \nonumber
\end{eqnarray}
Here $f$ denotes the semi-classical non-equilibrium phase-space distribution
of quasi-particles.
$m(\vec{x},t)$ denotes the sum of the bare and the dynamical
(space time dependent) mass generated by the Hartree term.
Within the spirit of kinetic theory one easily realizes
that the result obtained in (\ref{delpartb}) simply states that the change in the
occupation number per time $T$ is nothing but the collision rate. 
Within this Markovian approximation the fully dressed propagators are given by
\cite{GL98}
\begin{eqnarray}
\label{Dfullret}
D^{\rm ret}(\vec{x},t;p) & \approx &
\frac{1}{p^2 - m^2(\vec{x},t) - {\rm Re}\Sigma (\vec{x},t;p)
+ip_0 \Gamma (\vec{x},t;p)} \, ,
\\
\label{Dfullav}
D^{\rm av}(\vec{x},t;p) & \approx &
\frac{1}{p^2 - m^2(\vec{x},t) - {\rm Re}\Sigma (\vec{x},t;p)
- i p_0 \Gamma (\vec{x},t;p) } \, ,
\\
\label{Dfull<}
D^{<}(\vec{x},t;p) & \approx &
(-2i)
\frac{p_0 \Gamma (\vec{x},t;p)}{(p^2-m^2(\vec{x},t) -
{\rm Re} \Sigma (\vec{x},t;p) )^2 + p_0^2 \Gamma ^2(\vec{x},t;p)  }
\, f(\vec{x},t;\vec{p}) \, \, \, .
\end{eqnarray}
In particular we emphasize that
in (\ref{Dfull<}) the {\em instantaneous}
non-equilibrium phase space distribution function
$f(t)$ enters and not the initial one, $\tilde n$. 
The dynamically generated mass as well as the
collisional self energy contribution $\Sigma $ can thus be evaluated
with these propagators. (In kinetic theory one usually takes the propagators
in their quasi-free limit (${\rm Re}\Sigma$, $\Gamma$ $\to 0$), albeit 
instantaneous.)
Higher order terms leading to the pinch singularities
are explicitly resummed and lead now to finite and very transparent results. 

One can now easily understand the observations made by Altherr \cite{A95}.
He has found, starting from some non-equilibrium distribution $\tilde{n}$,
that higher order diagrams contribute to the same order
in the coupling constant as the lowest order one. Indeed, in his investigation,
the particular higher order diagrams where
nothing but the perturbative contributions
of the series in (\ref{DSD<1}) for the dressed
or resummed one-particle propagator $D^<$.
The only difference is that he has employed a `free' propagator
modified by some finite width in order that each of the terms in the series
(\ref{DSD<1}) becomes well defined.
The reason
for the higher order diagrams to contribute to the same order
is that the initial out-of-equilibrium distribution
$\tilde{n}$ cannot stay constant during the evolution of the system as it has
to evolve towards the Bose distribution.
If $\tilde{n} - n_B$ is of order $o(1)$,
it is obvious that there must exist contributions which perturbatively attribute 
to the
temporal change of the distribution function and contribute to the same order
$o(1)$.
In fact, in our prescription (\ref{Dfull<}),
$\tilde{n}$ has simply be substituted by the actual phase space distribution
$f$. Then calculating e.g. the tadpole diagram,
as discussed in the particular case of \cite{A95},
one has to stay within lowest order in the skeleton expansion,
but with the fully dressed propagator.

In summary, we have shown in simple physical terms
why so called pinch singularities do (and have to) appear in the
perturbative evaluation of higher order diagrams within the CTP description
of non-equilibrium quantum field theory. They are simply connected to
the standard divergence in elementary scattering theory.
The occurrence of pinch singularities signals the occurrence of (onshell) damping or
dissipation. This necessitates in the description of the evolution of
the system by means of non-perturbative transport equations.
In the weak coupling regime
this corresponds to standard kinetic theory. In this case we have given a 
prescription of how the dressed propagators can be approximated in a very 
transparent form. Technically, pinch singularities appear due to a misuse of Fourier
techniques \cite{B95}. From a physical point of view, scattering processes which
change the occupation number give rise to pinch singularities, if these processes
go on for infinitely long time. However, exactly these processes drive the system 
towards thermal equilibrium within a {\it finite} time characterized by the inverse
damping rate. In equilibrium the occupation number stays constant and no pinch
singularities can appear.

\acknowledgments
We gratefully acknowledge discussions with M.~Thoma.

\end{document}